\documentclass[preprints,accept,article,pdftex]{Definitions/mdpi} 
\RequirePackage{fix-cm}
\usepackage[T1]{fontenc}
\usepackage[utf8]{inputenc}

\usepackage{bm}
\usepackage{listings}
\usepackage{subfigure}
\usepackage{amsmath}
\usepackage{amssymb}
\usepackage{booktabs}
\usepackage{mathptmx}
\usepackage{enumitem}
\usepackage{hyperref}
\usepackage{siunitx}
\hypersetup{colorlinks,citecolor=blue}

\firstpage{1} 
\pubvolume{xx}
\issuenum{1}
\articlenumber{5}
\pubyear{2025}
\copyrightyear{2025}
\history{Received: Aug 15 2025}
\Title{Revisiting Stochastic Collocation with Exponential Splines for an Arbitrage-Free Interpolation of Option Prices}

\Author{Fabien {Le Floc'h}\orcidA{}}

\AuthorNames{Fabien {Le Floc'h}}

\address{%
	 Independent researcher}

\corres{Correspondence: {fabien@2ipi.com}}	
	\abstract{We revisit the stochastic collocation method using the exponential of a quadratic spline. In particular, we look in details whether it is more appropriate to fix the ordinates and optimize the abscissae of an interpolating spline or to fix the abscissae and optimize the parameters of a B-spline representation.}
	\keyword{stochastic collocation; implied volatility; quantitative finance; arbitrage-free; risk neutral density; B-spline}

\begin{document}

\section{Introduction}
\citet{lefloch2019model} introduce the stochastic collocation method utilizing either a spline or the exponential of a spline to achieve an arbitrage-free interpolation of vanilla option prices, or, equivalently, their implied volatilities. The formula for the undiscounted  price of a  call option $C$ with strike $K$ with the exponential spline collocation reads
\begin{align}
C(K) &= \int_{-\infty}^{\infty} \max(e^{g(x)} -K, 0) \phi(x) dx\,,
\end{align}
where $g$ is a monotonic quadratic spline function interpolating $(x_i, \ln y_i)_{i=0,...,N}$, $y_i$ are option strikes, $\phi$ is the normal probability density function.
 Let  $(\bar{x}_j)_{j=0,...,M}$ be the spline knots, and $g(x) = a_j + b_j(x-\bar{x}_j)+c_j(x-\bar{x}_j)^2$ on $[\bar{x}_j, \bar{x}_{j+1}]$, with $k$ the index such that $\bar{y}_k \leq K < \bar{y}_{k+1}$, we have then
\begin{align}
C(K) &= I(\tilde{x}_{-1},\bar{x}_0) + I(\tilde{x}_M,\infty) + K \Phi(-\tilde{x}_k)\nonumber\\
&+\sum_{j=k}^{M-1} e^{a_j - b_j \bar{x}_j + c_j \bar{x}_j^2 +\frac{1}{2}m_j^2}\frac{\Phi\left(\bar{x}_{j+1} \sqrt{1-2c_j}-m_j\right)-\Phi\left(\tilde{x}_j \sqrt{1-2c_j}-m_j\right)}{\sqrt{1-2c_j}} \,,
\end{align}
with $m_j=\frac{b_j-2c_j\bar{x}_j}{\sqrt{1-2c_j}}$ and defining $\tilde{x}_k =g^{-1}( \ln  K)$, $\tilde{x}_i = \bar{x}_i$ for $i >k$. We consider the cumulative normal density function in the complex plane: when $1-2c_j < 0$, we may use the imaginary error function erfi to compute its value.
For a linear extrapolation with slope $s$ and passing by the point $(x,y)$, corresponding to $g(u) = s (u-x) + y$, we find
\begin{align}
I(a,b)  &= \int_a^b e^{s(u-x)+y} \phi(u)du\nonumber\\
&= e^{y-s x + \frac{1}{2}s^2} \left( \Phi(b-s)  - \Phi(a-s)\right)\,,
\end{align}
with some abuse of notation $\Phi(-\infty) = 0$, $\Phi(\infty)=1$. The left wing extrapolation corresponds to $x=\bar{x}_0$ and $y={a}_0$, with $s=s_L$ a free parameter and the right wing extrapolation corresponds to $x=\bar{x}_M, y={a}_M$ with $s=s_R$.
In order to keep the continuity of the derivative at the boundaries, we choose $s_R=g'(x_M)$ and $s_L=g'(x_0)$.

The first moment is given by
\begin{align}
M_1 &=  \int_{-\infty}^{\infty} e^{g(x)} \phi(x)dx \nonumber\\
&=  I(-\infty,\bar{x}_0) +  I(\bar{x}_M,\infty) \nonumber\\
&+\sum_{j=0}^{M-1} e^{a_j - b_j \bar{x}_j + c_j \bar{x}_j^2 +\frac{1}{2}m_j^2}\frac{\Phi\left(\bar{x}_{j+1} \sqrt{1-2c_j}-m_j\right)-\Phi\left(\bar{x}_j \sqrt{1-2c_j}-m_j\right)}{\sqrt{1-2c_j}}\,.
\end{align}

In particular, \citet{lefloch2019model} explore the use of a B-spline parameterization with fixed knots
\begin{align}\label{eqn:bspline}
g(x) = \sum_{j=0}^{M} \alpha_j B_{j,3}(x)\,,
\end{align}
where the B-spline parameters $\alpha_j$ are calibrated such as to minimize the $\ell_2$-norm between weighted model prices and market prices. One challenge is to find a proper choice of knots in the $x$ space. Another approach, which is suggested but not fully explored is to consider a regular spline interpolation, fixing the ordinates $y$ instead of the knots $x$. The knots $x$ are calibrated and $y$ may be taken to be simply the market strikes. A standard B-spline parameterization is then not practical, since it would require solving a quadratic programming problem at each iteration of the non-linear least-squares minimization in order to enforce positivity of the B-spline coefficients, and thus monotonicity (see \citet{wolberg2002energy}). Instead, it is easier to rely on a monotonic spline interpolation representation. For quadratic splines, \citet{schumaker1983shape} proposes an algorithm to insert knots such that the interpolant preserves monotonicity and convexity and the spline is of class $C^1$. 

This latter approach has been rediscovered by \citet{roos2025simple}. Motivated by its encouraging results, we study in this note  the choice of fixing $y$ versus fixing $x$ in more details.

\section{When the Schumaker spline does not preserve monotonicity}
The quadratic spline from \citet{schumaker1983shape} preserves the shape, but not necessarily the monotonicity as \citep[Property 3.2]{schumaker1983shape} is not guaranteed to hold. A counter-example presented in Table \ref{tbl:schumaker_counter} and corresponds to a calibration on the market data of Table \ref{tbl:jackel} (case II). 
\begin{table}[h]
	\caption{Counter-example where the Schumaker spline $g$ on $(x,y)$ is not monotonic but the data is. We have $g'(3.03) < -0.0472 $.\label{tbl:schumaker_counter}}
\centering{\begin{tabular}{ll}\toprule
	x & y\\ \midrule
 -3.8732183006023453 &-3.34887695753723 \\
 -3.0522883128452993 & -3.0139892617835105 \\
 -1.7943713634054417 & -2.6791015660297828 \\
 -1.6512340998496167 & -2.344213870276071 \\
 -1.6259385165727456 & -2.009326174522346 \\
 -1.6006430286358648 & -1.6744384787686222 \\
 -1.578610520539529  & -1.3395507830148954 \\
 -1.5563726083805842 & -1.0046630872611706 \\
 -1.4706329281117003 & -0.6697753915074481 \\
 -1.3853673387319743 & -0.33488769575372446 \\
 -1.2808142988169438 &  0.0 \\
 -0.974725079691088  &  0.3348876957537252 \\
 -0.799531535567272  &  0.6697753915074496 \\
 -0.2964451400432737 &  1.00466308726117 \\
  0.40549865892657655&  1.3395507830148956 \\
  0.8534594889549447 &  1.6744384787686188 \\
  2.5550732737844375 &  2.009326174522343 \\
  2.822138011027902  &  2.3442138702760635 \\
  2.887184595942125  &  2.6791015660297917 \\
  3.1735718560441963 &  3.0139892617835136 \\
  3.294759066619504  &  3.348876957537241 \\ \bottomrule
\end{tabular}}
\end{table}

A solution is to use a harmonic mean estimate for the derivatives at the initial knots \citep{lam1990monotone}. This is the variation on the original Schumaker algorithm that we make use for our numerical experiments.

\section{The issue with fixing the ordinates}
\citet{lefloch2019model} pay attention to preserve the first moment exactly in order to guarantee the arbitrage-free property of the interpolation over the full range of strikes. Instead, the earlier stochastic collocation on a polynomial of \citet{grzelak2017arbitrage} was computing the call option price and deducing the put option price from the put-call parity relationship. In this latter approach, there is no guarantee then that the put option price will be positive for all strikes as the undiscounted call option price obtained by such a polynomial collocation may be below the intrinsic value $F-K$.

For a positive asset, if we assume absorption at zero, the put-call parity must be adjusted using the call option price at strike zero: $\tilde{F}=C(0)$. As the forward price comes from the market, the value of $\tilde{F}$ is given, and we are back to enforcing the preservation of the first moment.

The simplest way to enforce the first moment condition is to shift the ordinates of the interpolation by $\ln F - \ln C(0)$. This means that the ordinates are no longer truly fixed: at each iteration, the first moment condition will shift the ordinates. It is not obvious a priori if the shift is large or small.

In this analysis, the B-spline is always calibrated using fixed knots (abscissae) by optimizing its parameters, and the Schumaker spline is always calibrated from a fixed \emph{initial} set of ordinates (but which moves to enforce the first moment condition) by optimizing its abscissae.

We consider the example market data of \citet{jackel2014clamping} presented in Table \ref{tbl:jackel} (case II), and use two different initial guesses, with \emph{initial} ordinates at the market strikes: \begin{enumerate}[label=(\roman*)]
	\item The equivalent Black-Scholes implied volatility is constant equal to the at-the-money volatility.
	\begin{align}
	a_j = \ln K_j\,,\quad b_j = \sigma(F(T),T)\sqrt{T}\,,\quad x_j = \frac{1}{b_j}\left(a_j + \frac{b_j^2}{2} - \ln F(T)\right)\,,\label{eqn:abscissae_atm}
	\end{align}
	where $\sigma(K,T)$ is the market\footnote{The at-the-money volatility may computed by interpolating with a quadratic on the nearest neighbours.} Black-Scholes implied volatility at strike $K$ for maturity $T$.
	\item The abscissae and slopes use the full implied volatility smile.
 	\begin{align}
	a_j = \ln K_j\,,\quad b_j = \sigma({K_j,T})\sqrt{T}\,,\quad x_j = \frac{1}{b_j}\left(a_j + \frac{b_j^2}{2} - \ln F(T)\right)\,.\label{eqn:abscissae_smile}
	\end{align}
\end{enumerate}
In (i), the first moment is actually preserved directly and the initial ordinates correspond exactly to the market strikes. In (ii), a small shift of around 0.18 is incurred, which is well below the ordinate of next strike for this data (the spacing in log-strikes is approximately 0.33).
During the Levenberg-Marquardt minimization starting from the initial guess (i), the ordinates move significantly off the original strikes. This is not as much pronounced with initial guess (ii), but there is no guarantee that this will hold for other market data. 
\begin{figure}[H]
	\centering{
		\subfigure[Schumaker\label{fig:first_ordinate_expspline_jaeckel2}]{
			\includegraphics[width=.48\textwidth]{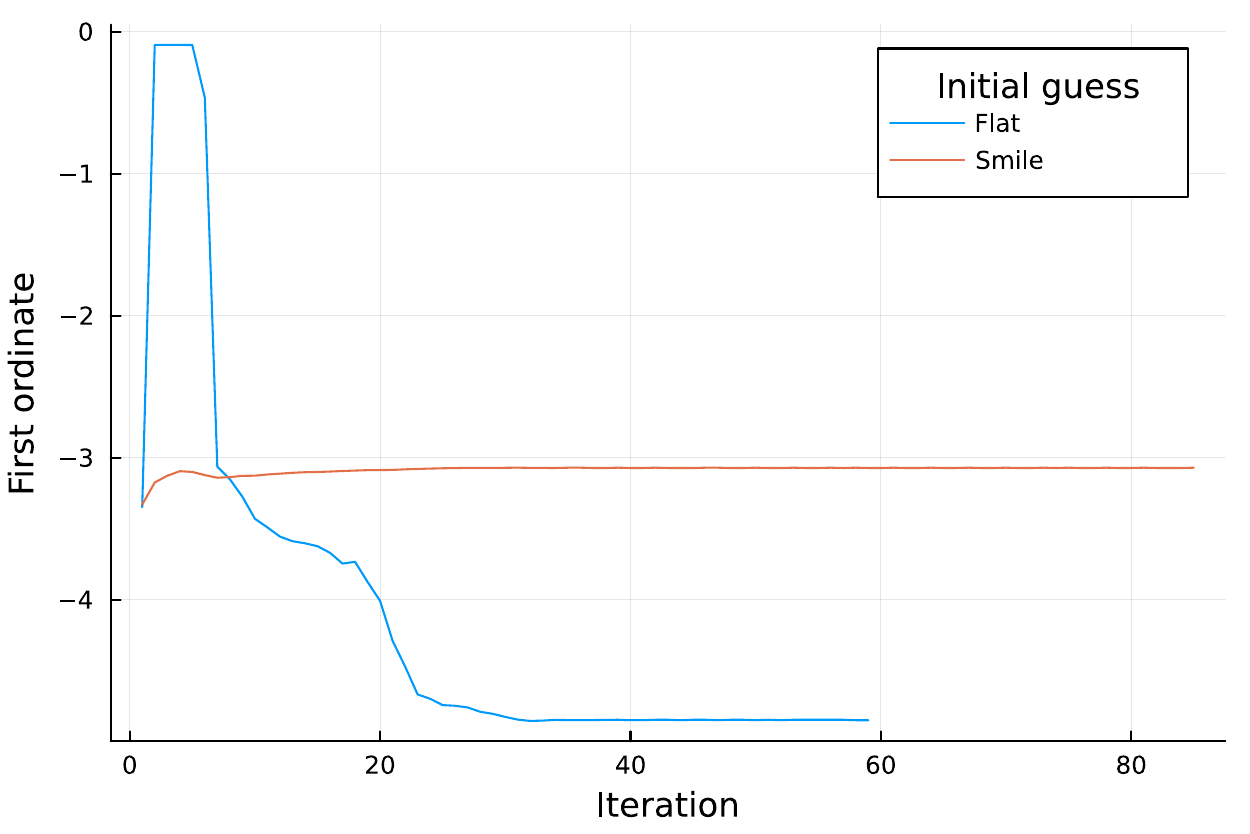}}
		\subfigure[B-spline\label{fig:first_ordinate_expbspline_jaeckel2}]{
			\includegraphics[width=.48\textwidth]{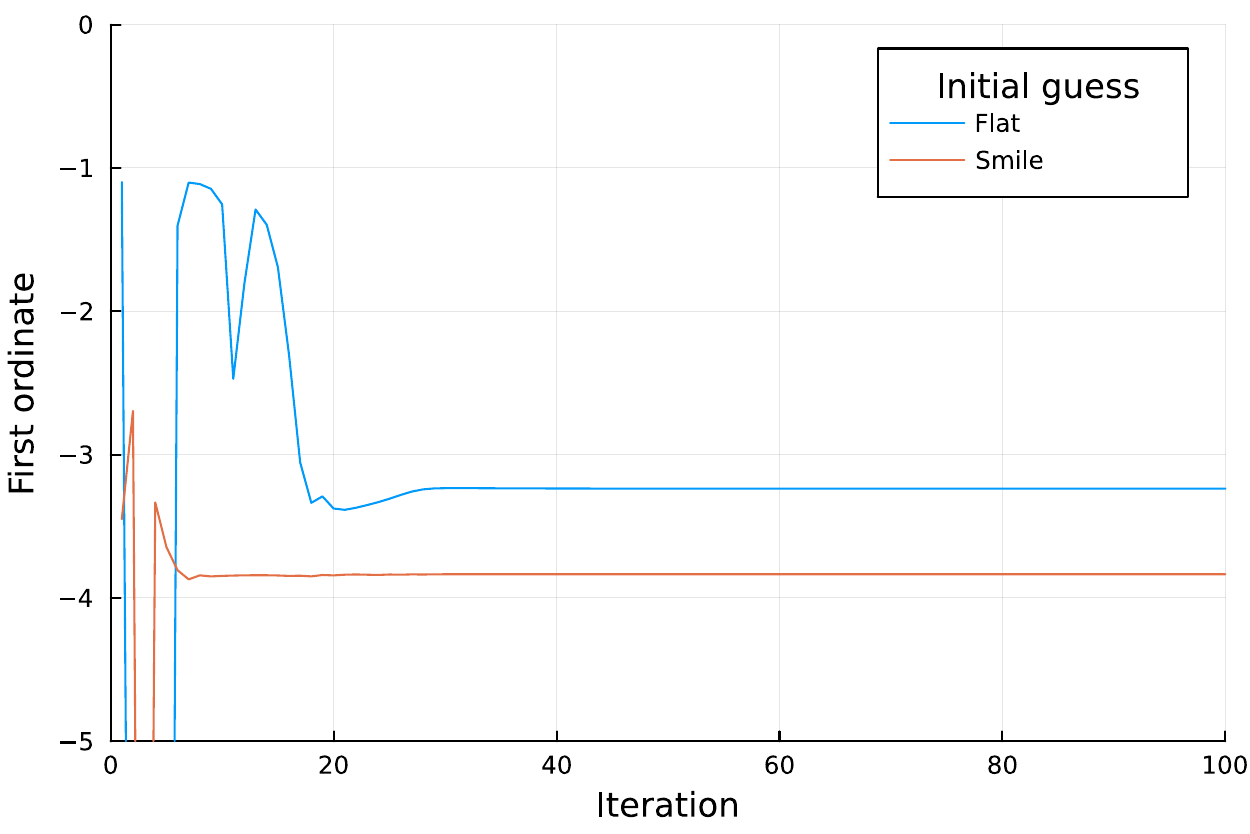} }
	}
	\caption{First ordinate of the exponential spline collocation as a function of the iteration of the Levenberg-Marquardt minimizer. The minimizer is applied to the abscissae for the Schumaker spline and to the coefficients for the B-spline.}
\end{figure}

Figure \ref{fig:first_ordinate_expbspline_jaeckel2} presents the variation of the lowest ordinate during the minimization when using the exponential B-spline collocation of \citet{lefloch2019model}, with the two different kinds of initial guess. Those are not exactly the same as previously since the abscissae $x$ is computed using the guidelines of \citet{lefloch2019model} from the estimated cumulative probability density and thus the initial ordinates $a_j$ do not correspond exactly to the market option strikes. The relations between $x_j$, $a_j$ and $b_j$ are however the same. We truncated the graph to the same limits as Figure \ref{fig:first_ordinate_expspline_jaeckel2}, otherwise the minimum ordinate goes below -27 and the number of iterations above 750.

\begin{figure}[H]
	\centering{
		\subfigure[Schumaker\label{fig:iv_expspline_jaeckel2}]{
			\includegraphics[width=.48\textwidth]{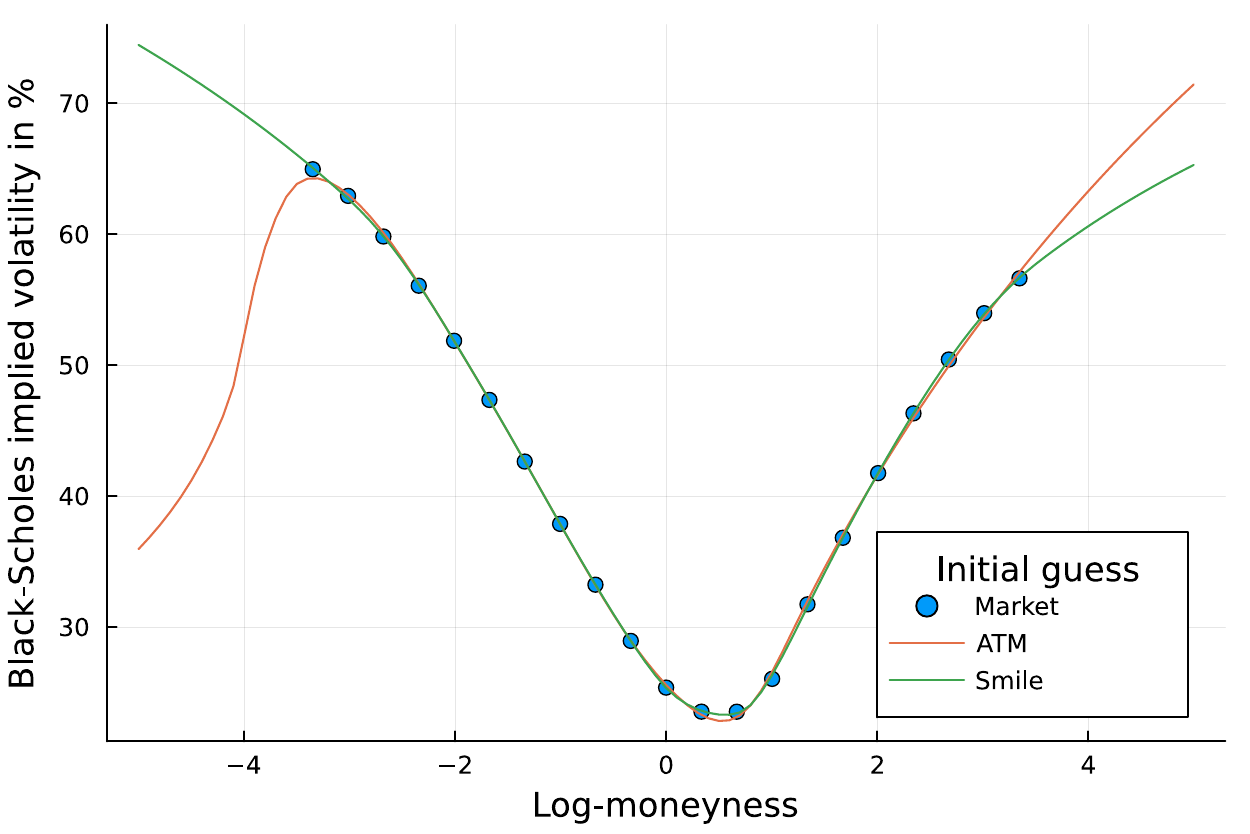}}
		\subfigure[B-spline\label{fig:iv_expbspline_jaeckel2}]{
			\includegraphics[width=.48\textwidth]{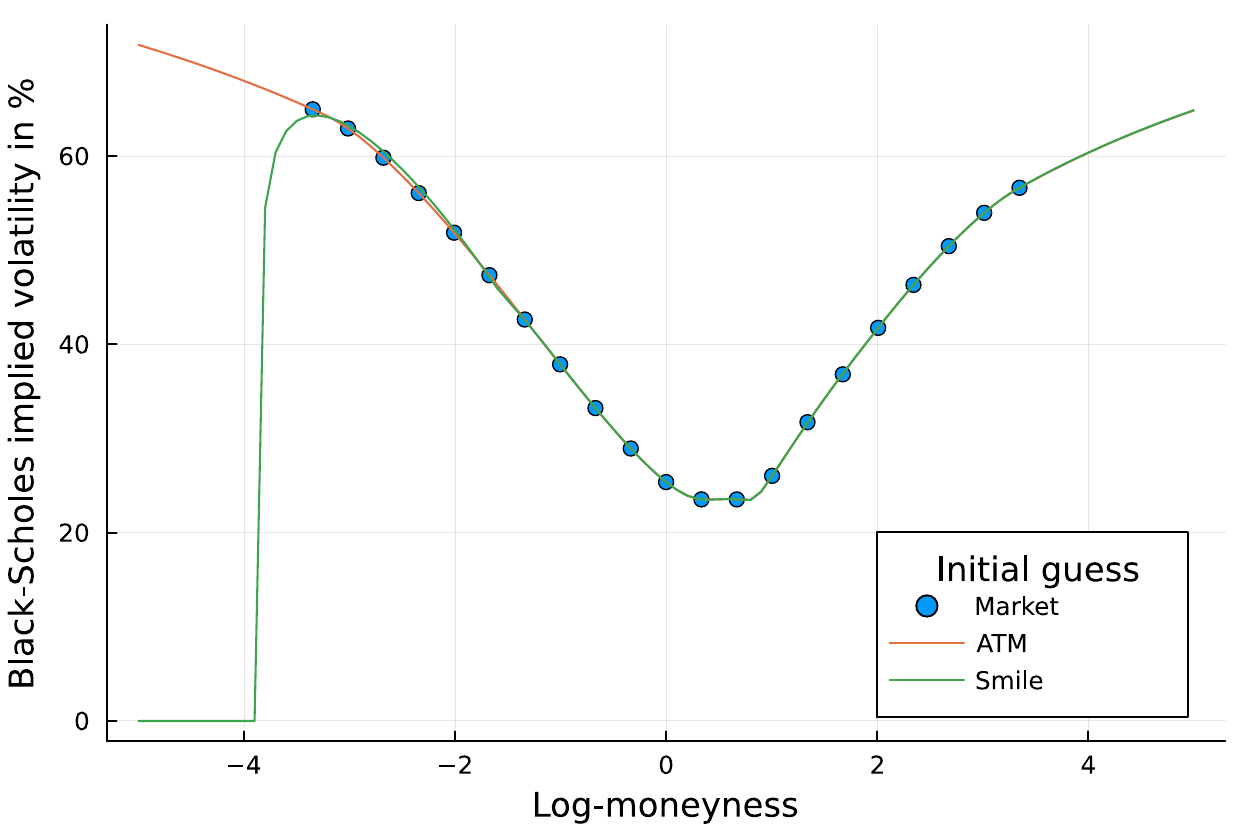} }
	}
	\caption{Implied volatility of the exponential spline collocation calibrated starting from two different initial guesses.}
\end{figure}

With both parameterization, the initial guess (i) leads to a different solution than initial guess (ii). And even though we "fixed" the ordinates of the Schumaker spline, the end solutions possess significantly different ordinates, with initial guess (ii) being much closer to the market strikes and leading to the best fit in terms of implied volatilities (Figure \ref{fig:iv_expspline_jaeckel2}). 

Interestingly, the implied volatility smile resulting from initial guess (i) and the optimization on the abscissae exhibits an awkward left wing. This also happens with initial guess (ii) and the optimization on the B-spline parameters. We have observed this behavior on occasion when calibrating the exponential spline collocation model without regularization, when the first ordinate is too far off the lowest strike.

\begin{figure}[H]
	\centering{
		\subfigure[First ordinate\label{fig:first_ordinate_expbspline_jaeckel2_eps0001}]{
			\includegraphics[width=.48\textwidth]{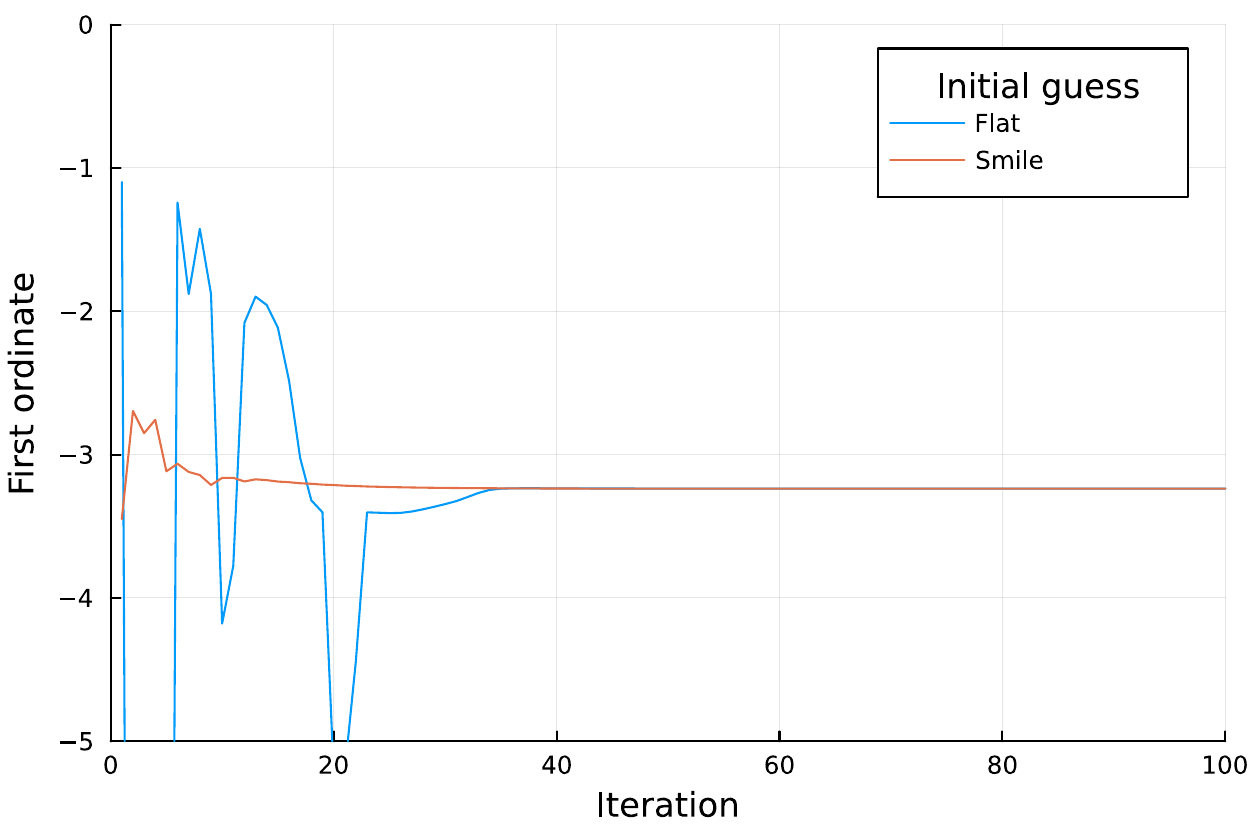}}
		\subfigure[Implied volatility\label{fig:iv_expbspline_jaeckel2_eps0001}]{
			\includegraphics[width=.48\textwidth]{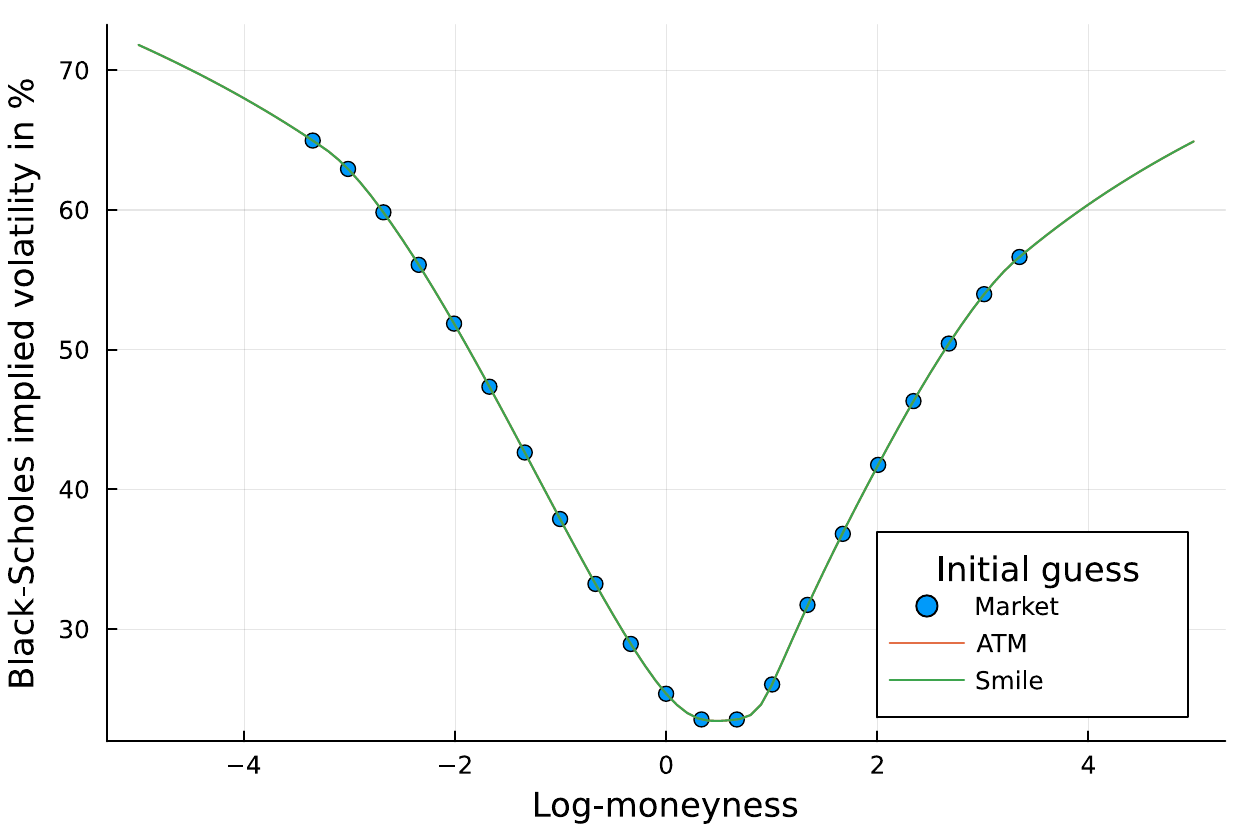} }
	}
	\caption{Calibration of the exponential B-spline collocation from two different initial guesses with regularization constant $\epsilon=10^{-4}$.}
\end{figure}

We found the penalization on the difference of inverse slopes to be effective\footnote{The penalty on the second-derivative values described in \citep{lefloch2019model} also works well on the examples considered here.} for the B-spline parameterization:
\begin{align}
	E_{\textmd{penalty}} = \epsilon^2 \sum_{j=0}^{M-1} \left[\frac{1}{b_{j+1}}-\frac{1}{b_j}\right]^2\,.\label{eqn:bspline_penalty}
\end{align}
This regularization also has the advantage of avoiding spurious spikes in the probability density due to the derivative of $g$ being close to zero. Furthermore, $b_j$ will not explode because the difference in B-spline parameters $\lambda_{j+1}-\lambda_{j}$ is limited to a fixed interval $[0, \lambda_{\max}]$ during the minimization. 
A very small regularization makes both initial guesses converge to the same smoother solution. In Figure \ref{fig:iv_expbspline_jaeckel2_eps0001}, we use $\epsilon=10^{-4}$, but this is already true with $\epsilon=10^{-6}$, and also works with $\epsilon=10^{-2}$.

 Unfortunately, this regularization is not directly applicable to the Schumaker spline, since it has a varying number of knots. One possibility may be to artificially insert knots such that the number of knots is kept constant, but then one has also to ensure that $b_j$ and $c_j$ do not explode during the minimization. More simply, we may enforce a penalty similar to Equation \ref{eqn:bspline_penalty} at the market strikes:
 \begin{align}
	E_{\textmd{penalty}} = \epsilon^2 \sum_{i=0}^{N-1} \left[\frac{1}{g'(g^{-1}(K_{i+1}))}-\frac{1}{g'(g^{-1}(K_i))}\right]^2\,.\label{eqn:spline_penalty}
\end{align}
While this is reasonably effective when calibrating to a dense set of noisy option prices, it however does not help the calibration starting from initial guess (i) in our example.

\section{TSLA example}
We consider options on the stock with ticker TSLA expiring on January 17, 2020, as of June 15, 2018. We first imply the forward price from the put-call parity relation at-the-money and then imply the Black-Scholes volatility from the mid price for each option strike (Table \ref{tbl:tsla_jun1518}). In this example, the options mid prices are not arbitrage free.

Without penalization, the implied probability density would be extremely noisy due to the over-fitting of the market prices, given that this non parametric approach is equivalent to one less parameter than the number of market prices (see \citet{lefloch2019model} for an illustrative example on the same market data). We start with a penalty of $\epsilon=10^{-2}$ which leads to a reasonably smooth implied probability density. The implied probability density $D$ at strike $K$ may be computed by a second-order differentiation of the call option price or more directly through
\begin{align}
\textmd{D}(K)= \frac{\phi(g^{-1}(K))}{K g'(g^{-1}(K))}\,,
\end{align}
where $\phi$ is the normal probability density function.
\begin{figure}[H]
	\centering{
		\subfigure[Schumaker\label{fig:density_expspline_tsla_eps01}]{
			\includegraphics[width=.48\textwidth]{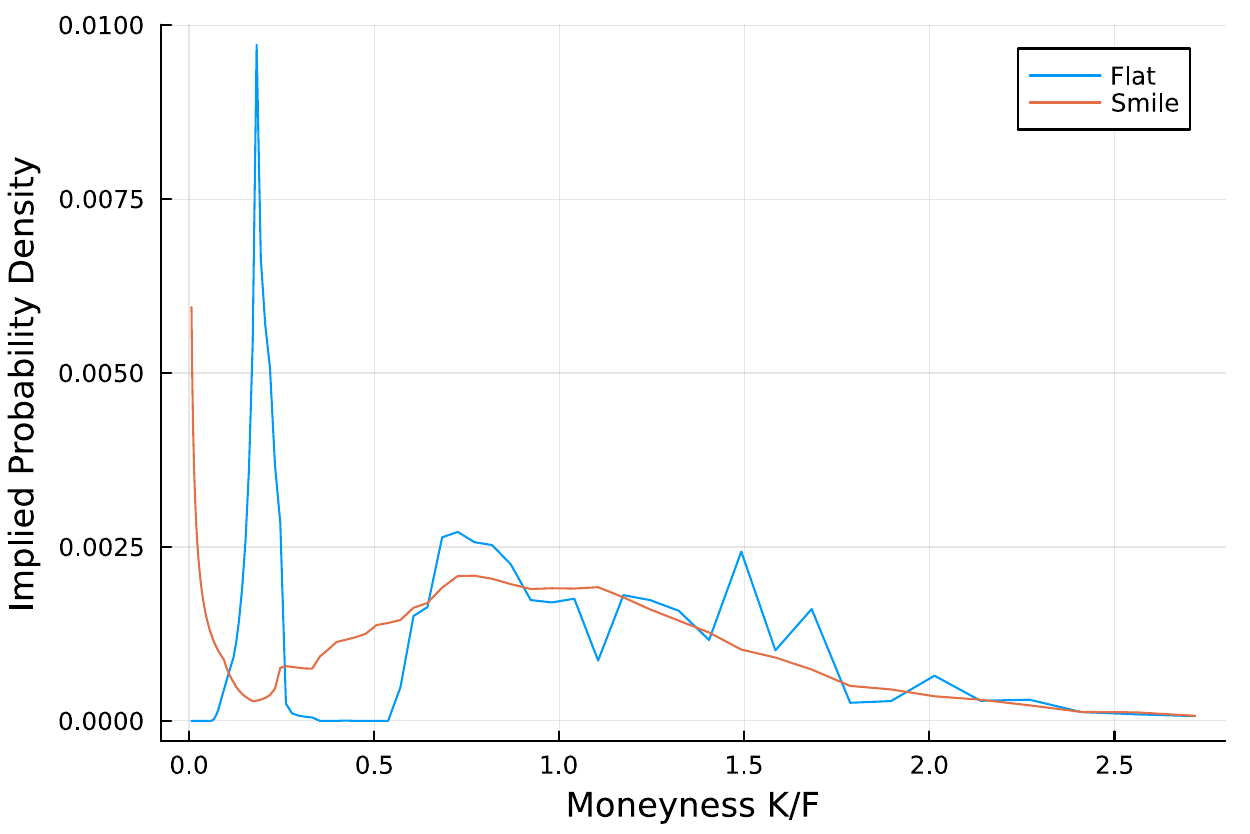}}
		\subfigure[B-spline\label{fig:density_expbspline_tsla_eps01}]{
			\includegraphics[width=.48\textwidth]{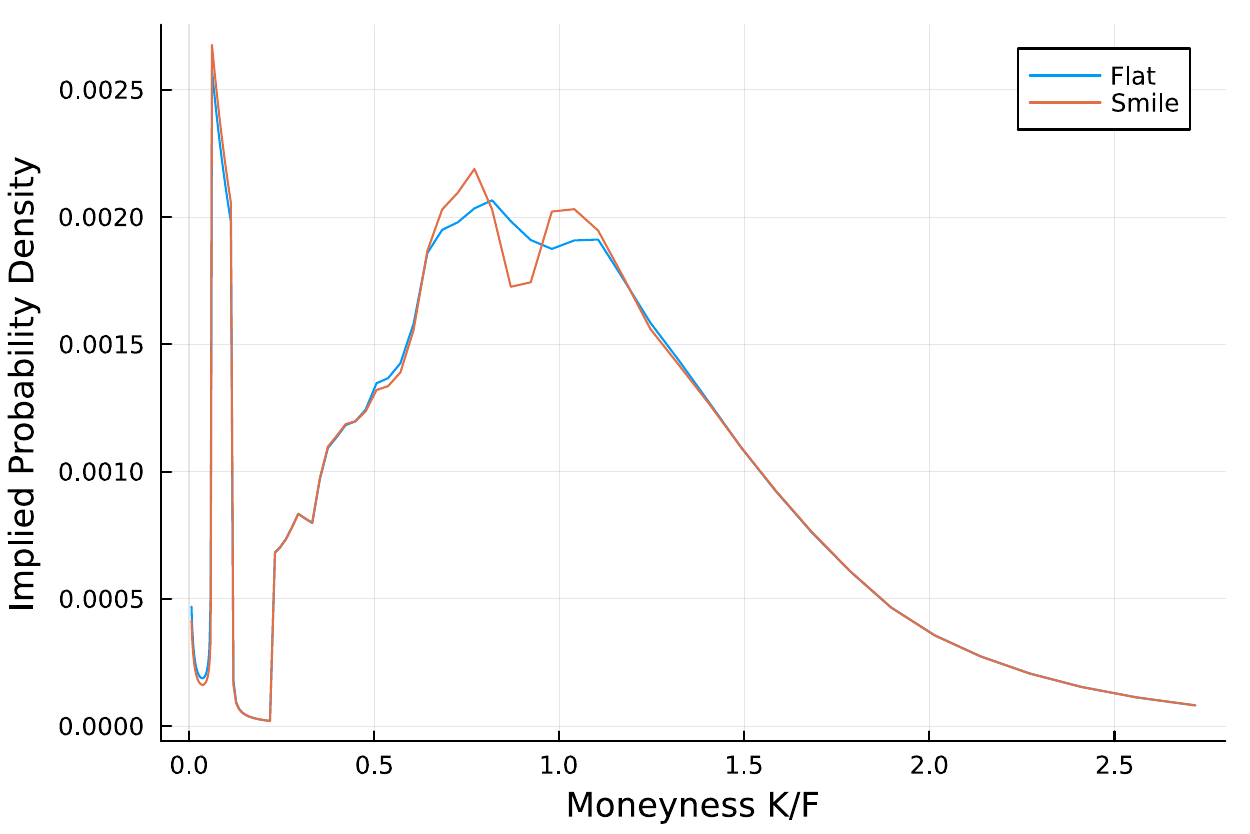} }
	}
	\caption{Implied probability density of the exponential spline collocation calibrated starting from two different initial guesses with regularization constant $\epsilon=10^{-2}$ on the TSLA options of maturity 18m.}
\end{figure}

\begin{figure}[H]
	\centering{
		\subfigure[Schumaker\label{fig:iv_expspline_tsla_eps01}]{
			\includegraphics[width=.48\textwidth]{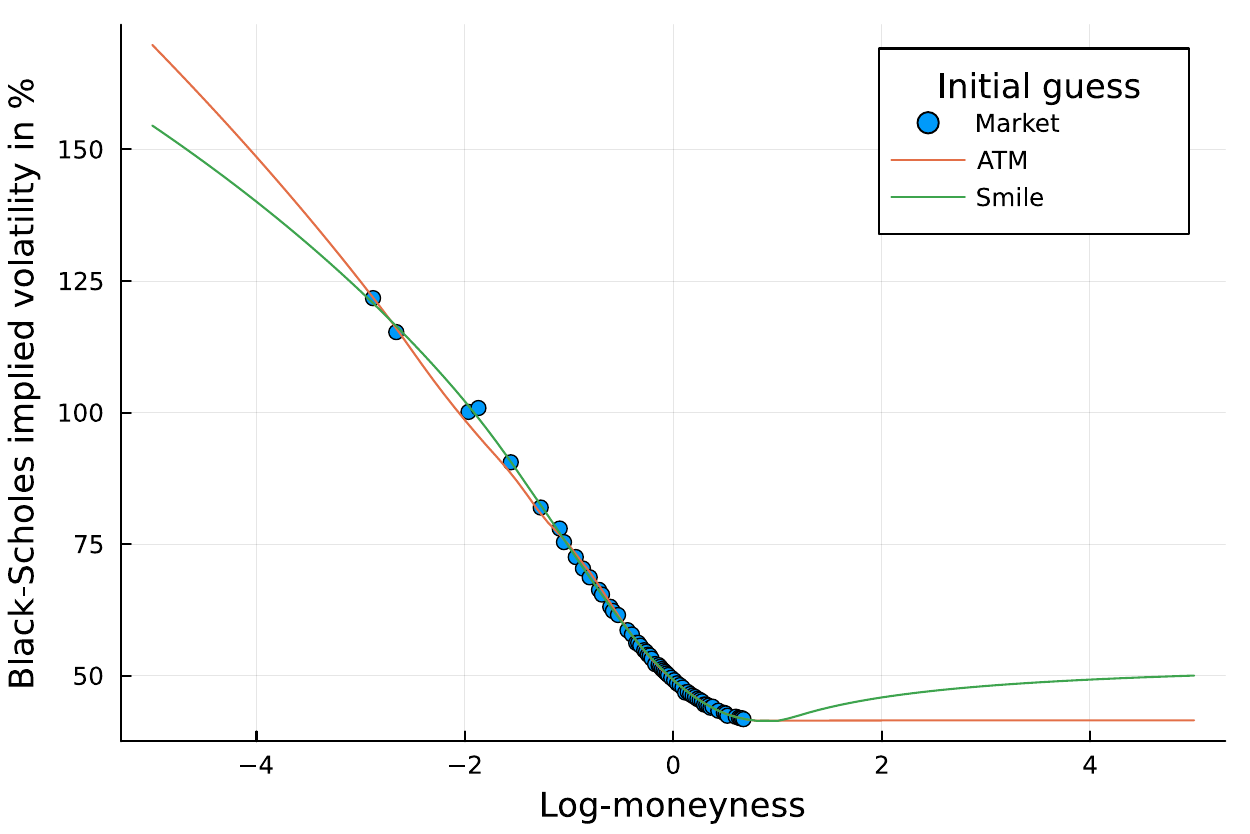}}
		\subfigure[B-spline\label{fig:iv_expbspline_tsla_eps01}]{
			\includegraphics[width=.48\textwidth]{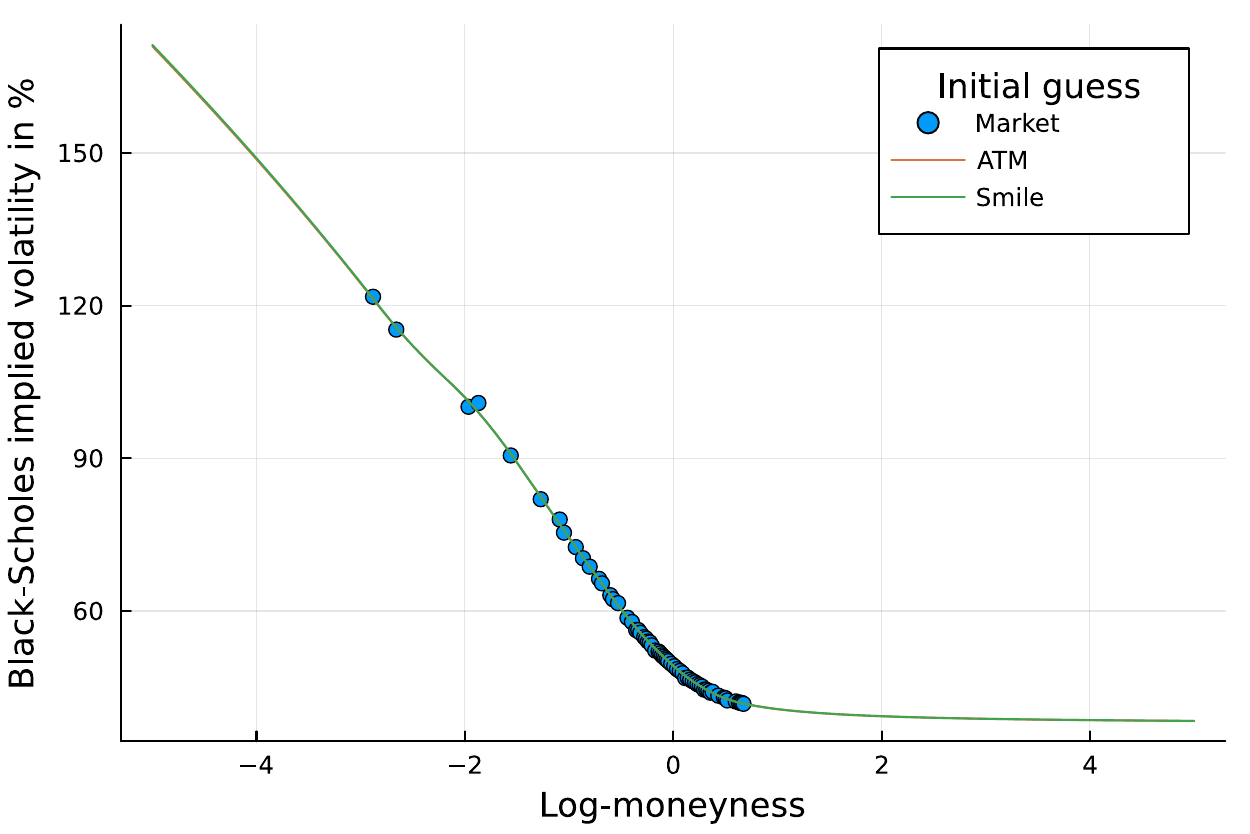} }
	}
	\caption{Implied volatility of the exponential spline collocation calibrated starting from two different initial guesses with regularization constant $\epsilon=10^{-2}$ on the TSLA options of maturity 18m.}
\end{figure}

\begin{figure}[H]
	\centering{
	\subfigure[Schumaker\label{fig:first_ordinate_expspline_tsla_eps01}]{
			\includegraphics[width=.48\textwidth]{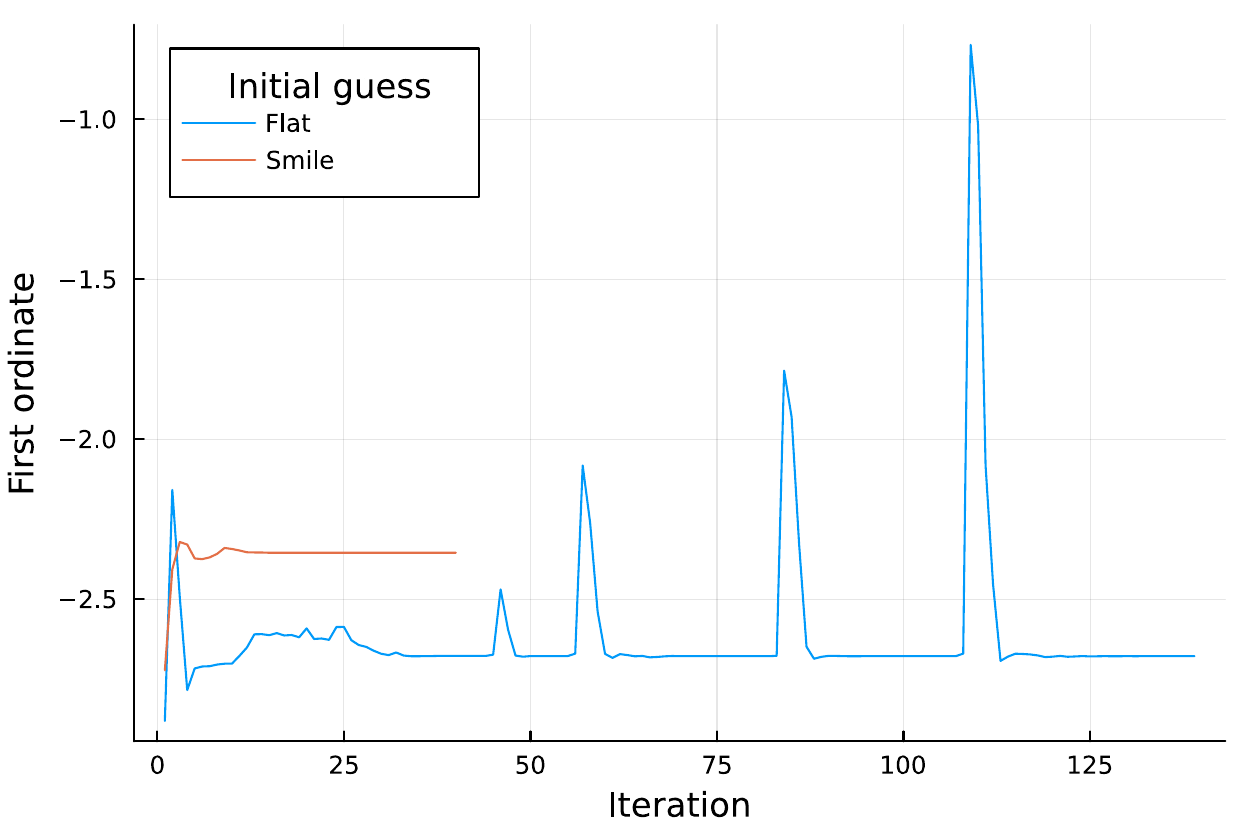}}
		\subfigure[B-spline\label{fig:first_ordinate_expbspline_tsla_eps01}]{
			\includegraphics[width=.48\textwidth]{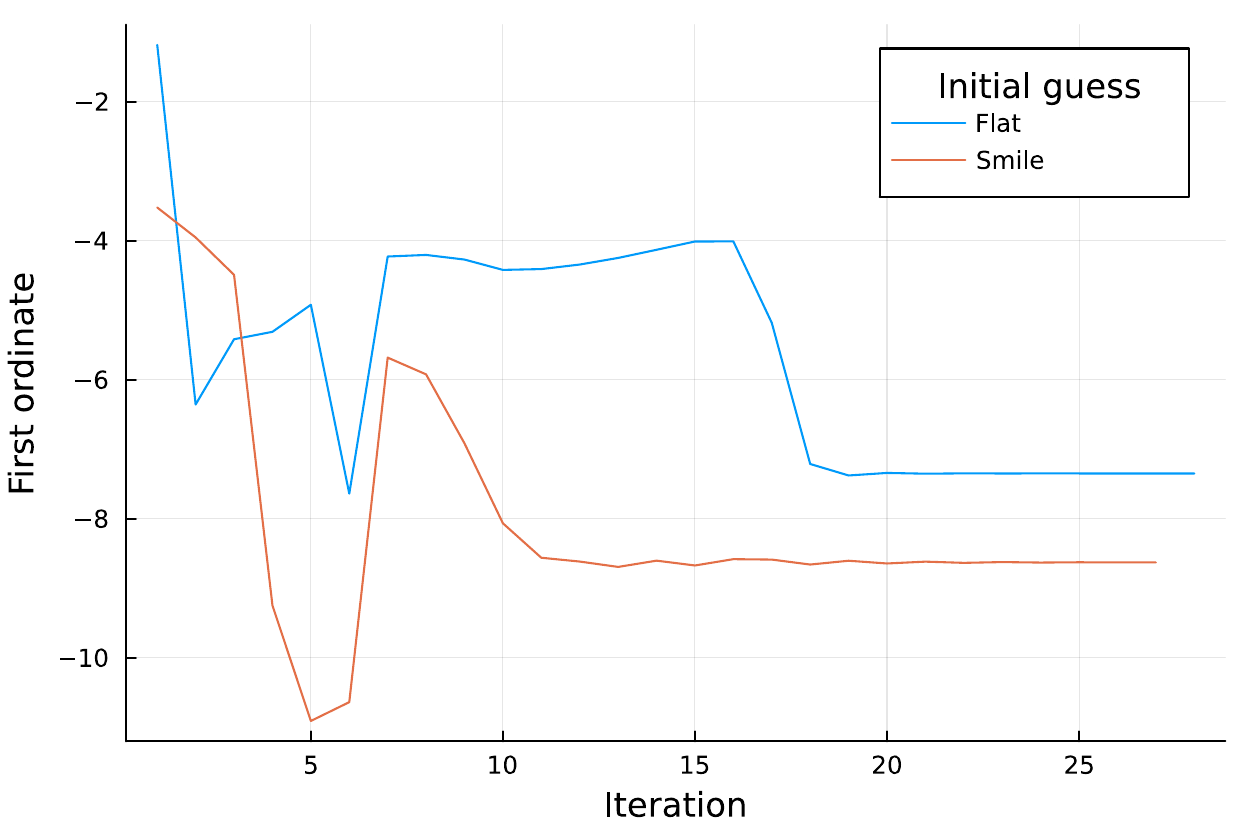} }
	}
	\caption{First ordinate of the exponential spline collocation as a function of the iteration of the Levenberg-Marquardt minimizer. The minimizer is applied to the abscissae for the Schumaker spline and to the coefficients for the B-spline  with regularization constant $\epsilon=10^{-2}$ on the TSLA options of maturity 18m.}
\end{figure}

The regularization works well for the B-spline with both initial guesses, and the Schumaker spline with initial guess (ii), but does not seem very effective on initial guess (i) and the Schumaker spline (Figure \ref{fig:density_expspline_tsla_eps01}). This is confirmed by the 
plot of the implied volatility smiles (Figure \ref{fig:iv_expspline_tsla_eps01}).
Initial guess (i), applied to the Schumaker spline, leads to not so great fit of the market options for low strikes. It is stuck in some suboptimal local minimum. The ordinates vary significantly during the minimization, even with initial guess (ii) (Figure \ref{fig:first_ordinate_expspline_tsla_eps01}). The market lowest strike is not so close to the optimal first ordinate anymore.

\begin{figure}[H]
	\centering{
		\subfigure[Schumaker\label{fig:iv_expspline_tsla_eps0001}]{
			\includegraphics[width=.48\textwidth]{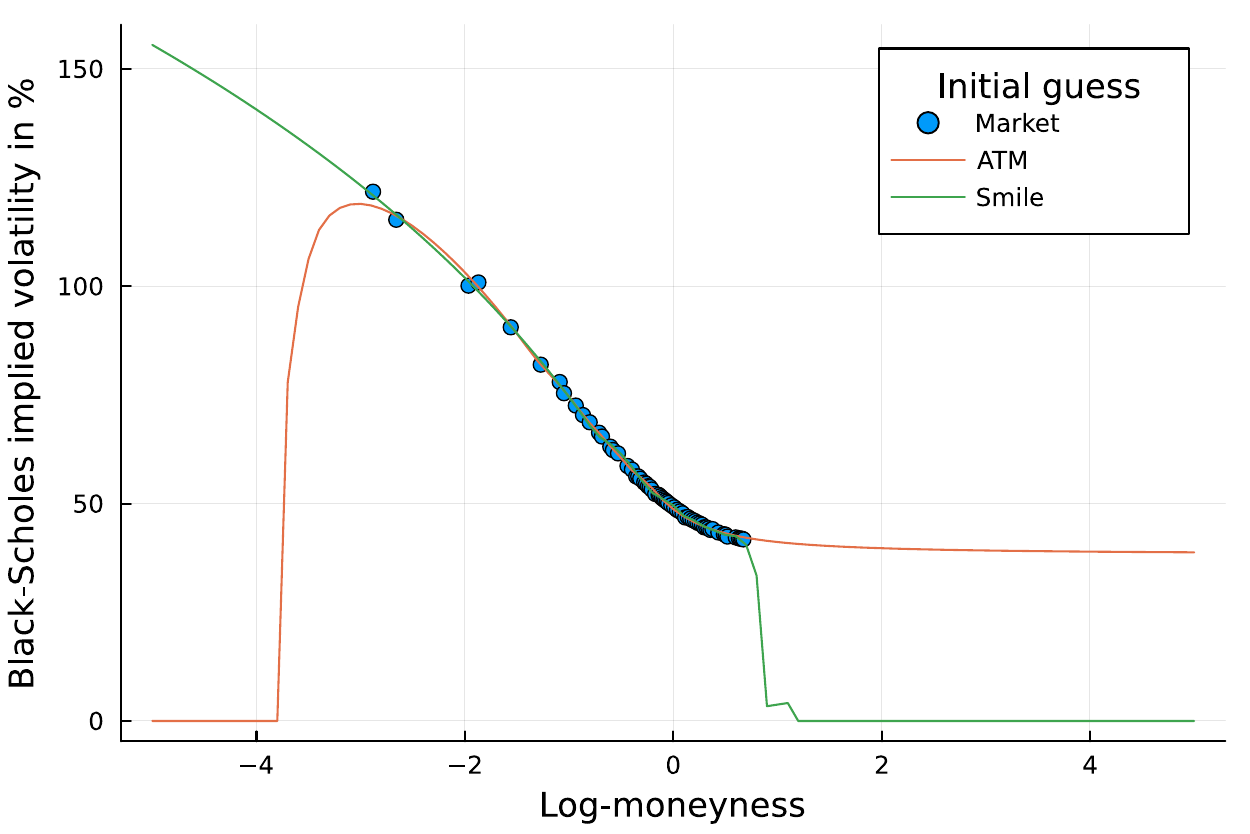}}
		\subfigure[B-spline\label{fig:iv_expbspline_tsla_eps0001}]{
			\includegraphics[width=.48\textwidth]{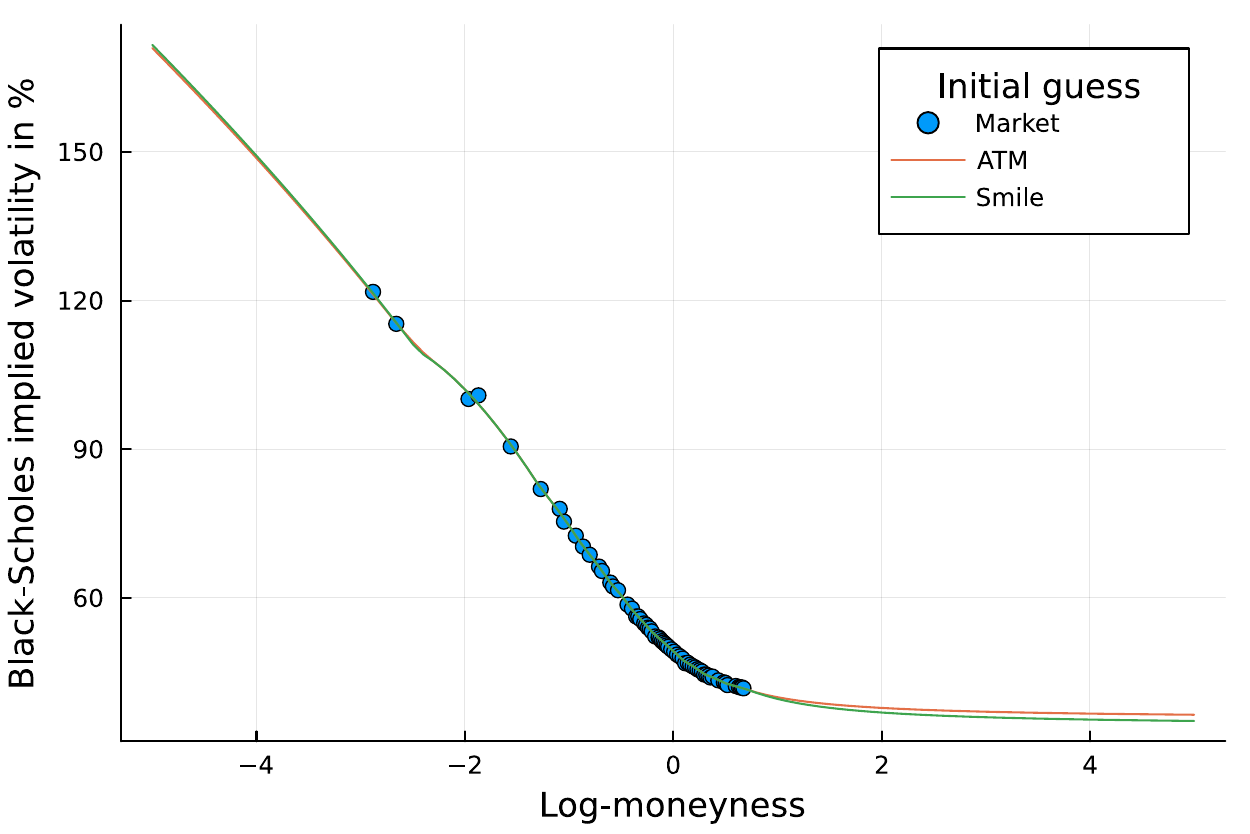} }
	}
	\caption{Implied volatility of the exponential spline collocation calibrated starting from two different initial guesses with regularization constant $\epsilon=10^{-4}$ on the TSLA options of maturity 18m.}
\end{figure}

It is enlightening to experiment with a smaller penalty of $\epsilon=10^{-4}$. The optimization of the abscissae becomes then more problematic. Even with the initial guess (ii), the Schumaker spline leads to a non-realistic smile for strikes just above the quoted market strikes (the prices go to zero in Figure \ref{fig:iv_expspline_tsla_eps0001}). One issue is that $x_M=98.93$, way too large\footnote{The odd shape and prices do not originate from the extrapolation part of the spline.}. As the ordinates are shifted upwards during the minimization, many $x_i$ do not directly correspond to market strikes anymore. One root cause is over-parameterization, and we did not impose a bound on $x_M$, only local bounds on each $x_i-x_{i-1}$. 
One way to impose both would be to increase the dimension of the unconstrained variable by one, which would correspond to a scaling variable. It is however not obvious that the resulting minimization would be more stable. 

With the B-spline parameterization, there is no such issue, and it leads to sensible smiles regardless of the penalty value. And again, both initial guesses lead to the same calibrated smile (Figure \ref{fig:iv_expbspline_tsla_eps0001}) showing the increased stability of the B-spline optimization compared to the Schumaker spline optimization.

\section{The issue with fixing the abscissae}
One concern with the B-spline approach, where the abscissae are fixed during the minimization, is the impact of the initial choice of knots on the outcome of the calibration. 

We consider the following  sets of knots:
\begin{enumerate}[label=(\roman*)]
\item follows the guidelines of \citet{lefloch2019model}
\item  uses the more direct x implied by market strikes as per Equation \ref{eqn:abscissae_smile}.
\item uses the more direct x implied by market strikes as per Equation \ref{eqn:abscissae_atm}.
\end{enumerate}

Table \ref{tbl:abscissae_bspline_jaeckel2} gives the abscissae $x_i$ for the market data corresponding to case II of Table \ref{tbl:jackel}.
\begin{table}[H]
	\caption{Abscissae $x_i$ of the B-spline for the three choices applied to case II of Table \ref{tbl:jackel}.\label{tbl:abscissae_bspline_jaeckel2}}
	\centering{
	\begin{tabular}{ccc}\toprule
		(i) & (ii) & (iii)\\\midrule
 -1.6364870807848217&-1.557018972731845    & -5.574166094776138\\ 
 -1.6261066909513329& -1.4176277692909673  & -4.988175014864264\\ 
 -1.5975883790170498& -1.3143421027079067  & -4.402183934952374\\ 
 -1.5581511199286542& -1.2247732224484822  & -3.8161928550405135\\ 
 -1.5022371676472155& -1.1359932381577464  & -3.230201775128629\\ 
 -1.420254328637289 & -1.0369270968395223  & -2.644210695216747\\
 -1.2974174526037123& -0.9145863522938157  & -2.0582196153048598\\ 
 -1.1130018044433514& -0.7510486016849476  & -1.4722285353929758\\ 
 -0.8408928570949368& -0.5205025891308158  & -0.8862374554810963\\ 
 -0.4514846087805624& -0.18789909427996598 & -0.30024637556921446\\ 
  0.0814689666759307&  0.28574470434266874 &  0.28574470434266874\\ 
  0.7305751589669239&  0.8967881631396702  &  0.8717357842545531\\ 
  1.4317407577681716&  1.5286684100766386  &  1.4577268641664363\\ 
  2.1819960086733534&  2.0063488588960663  &  2.0437179440783124\\ 
  3.1422990325335705&  2.2315836409702987  &  2.6297090239901975\\ 
  3.2272366371883887&  2.4338322594648223  &  3.2157001039020785\\ 
  3.422427747260656 &  2.6067615549820484  &  3.801691183813961\\
  3.4443005865798333&  2.7685989907187096  &  4.3876822637258375\\ 
  3.4453202609481752&  2.926428984110724   &  4.973673343637727\\ 
  3.4453469662892697&  3.0872115685336663  &  5.559664423549606\\ 
  3.4453473893808613&  3.263200609709557   &  6.1456555034614935\\ \bottomrule
	\end{tabular}}
\end{table}

\begin{figure}[H]
	\centering{
		\subfigure[Implied volatility]{
			\includegraphics[width=.48\textwidth]{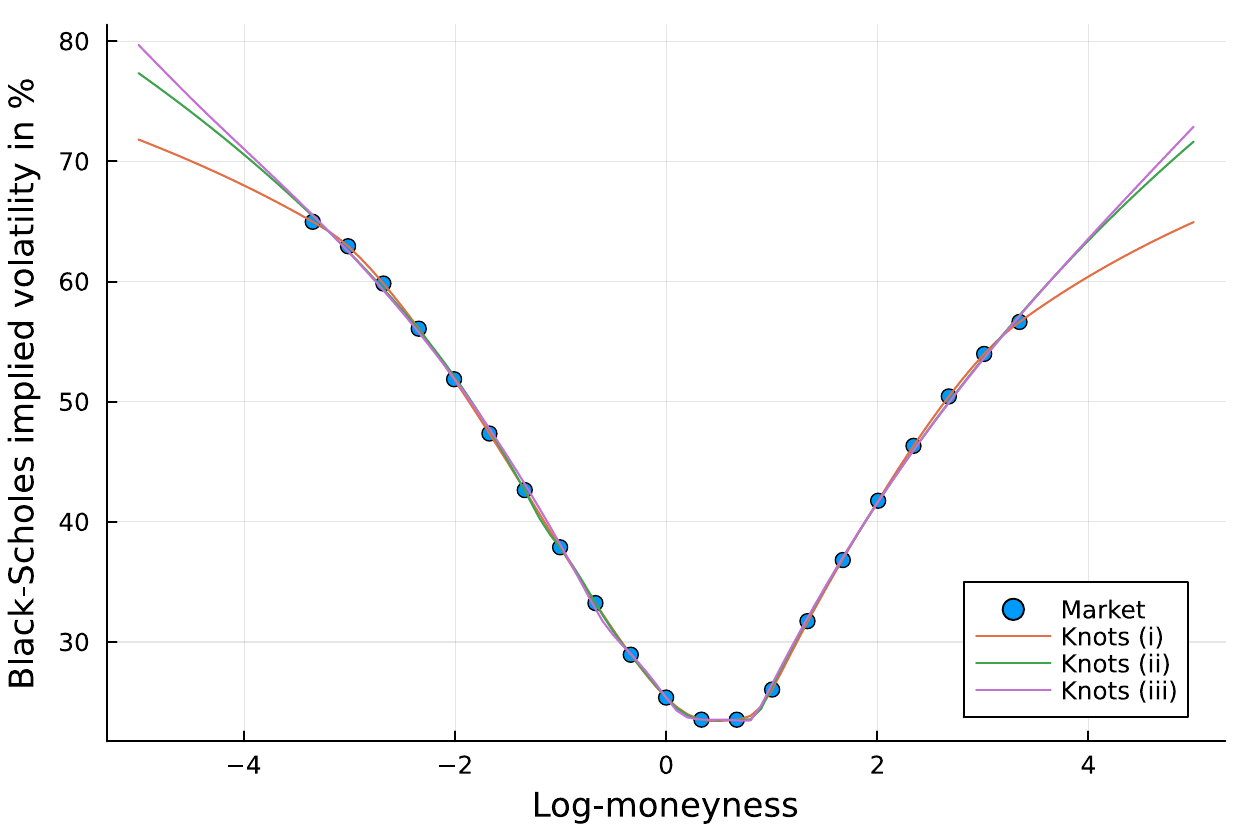}}
		\subfigure[Implied probability density\label{fig:density_expbsplinem_jackel_eps0001}]{
			\includegraphics[width=.48\textwidth]{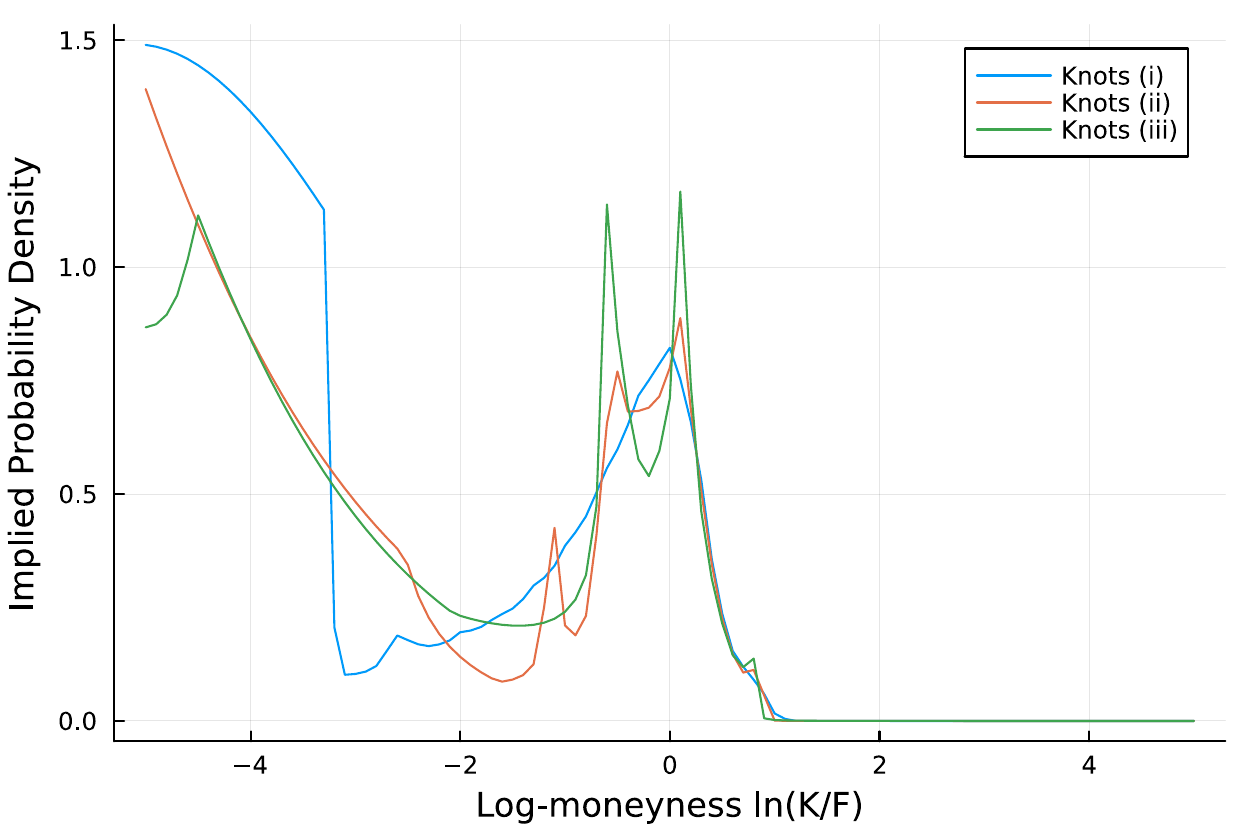} }
	}
	\caption{Implied volatility and probability density of the exponential B-spline collocation calibrated with different sets of knots, and regularization constant $\epsilon=10^{-4}$ on case II of Table \ref{tbl:jackel}.\label{fig:iv_expbsplinem_jackel_eps0001}}
\end{figure}

The calibration leads to a much better fit for the set (i). But the set (iii) leads to a similar smile and a similar implied probability density (Figure \ref{fig:iv_expbsplinem_jackel_eps0001}) with a penalty factor $\lambda=10^{-2}$. The implied density is quite different and somewhat awkward with knots (ii) or (iii) for smaller penalties such as $\lambda \leq 10^{-4}$ (Figure \ref{fig:density_expbsplinem_jackel_eps0001}).
\begin{figure}[H]
	\centering{
		\subfigure[Implied volatility]{
			\includegraphics[width=.48\textwidth]{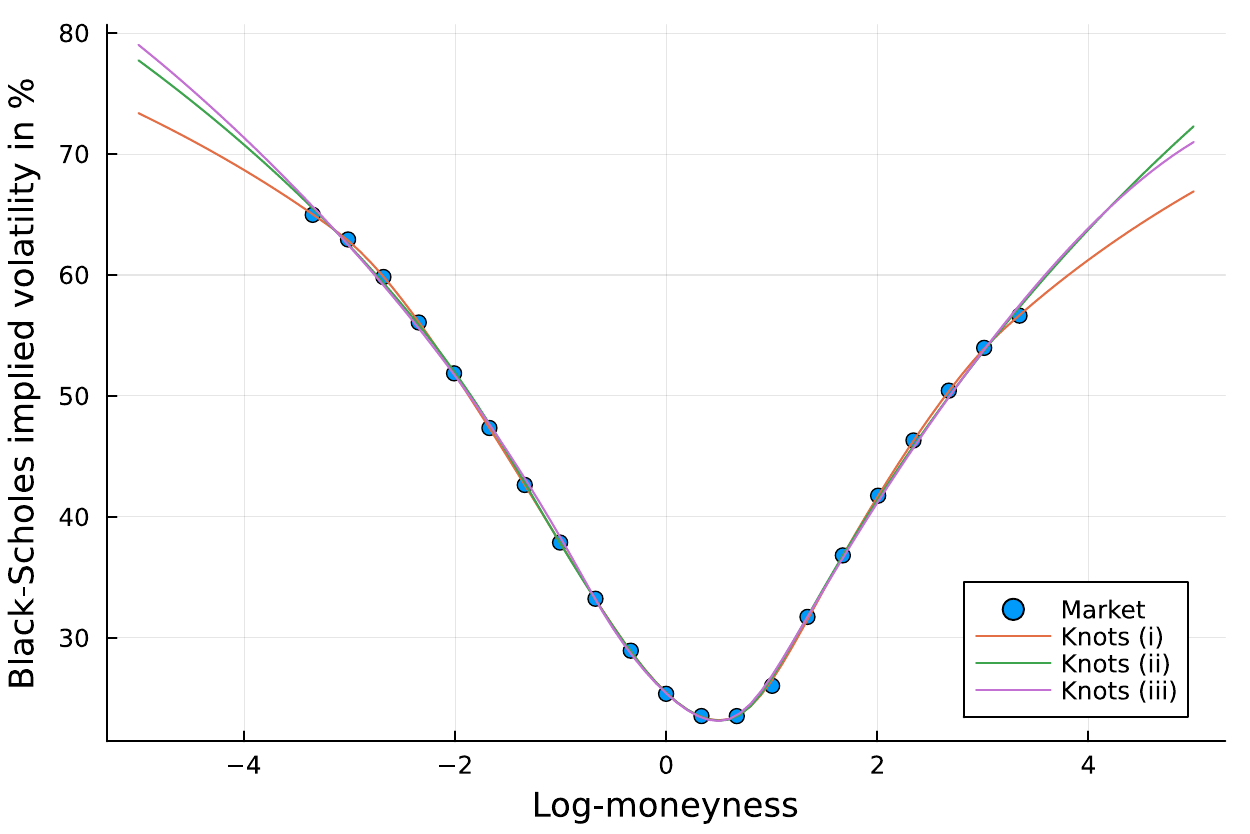}}
		\subfigure[Implied probability density\label{fig:density_expbsplinem_jackel_eps01}]{
			\includegraphics[width=.48\textwidth]{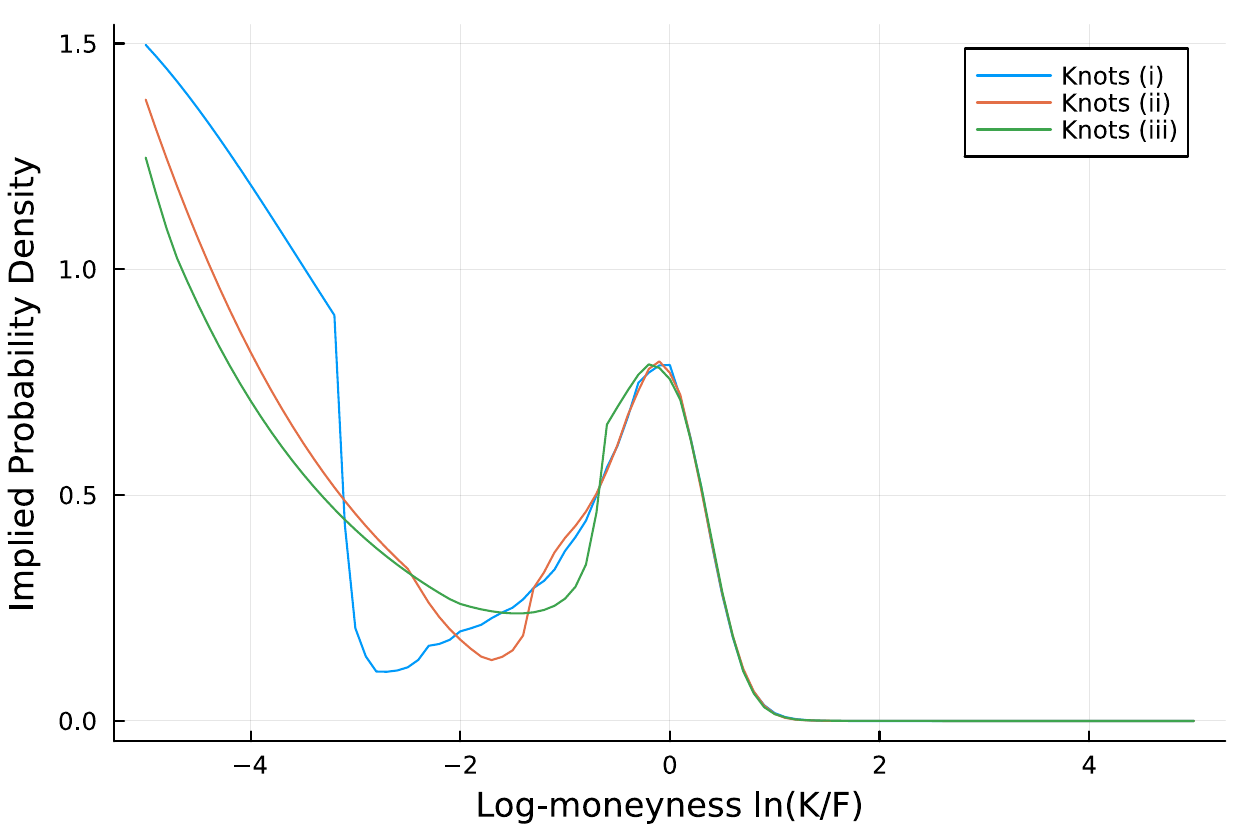} }
	}
	\caption{Implied volatility and probability density of the exponential B-spline collocation calibrated with different sets of knots, and regularization constant $\epsilon=10^{-2}$ on case II of Table \ref{tbl:jackel}.\label{fig:iv_expbsplinem_jackel_eps0001}}
\end{figure}

On the TSLA 18m options, the results are even closer between the different choices of knots.

\section{On the use of a quadratic exponential extrapolation}
\citet{roos2025simple} proves if we extrapolate with the exponential of a quadratic function,  the quadratic coefficient determines the number of finite moments and thus the slope of the implied variance as a function of log-moneyness.
For a left extrapolation defined by
\begin{align}
g(x) = c_{\textmd{left}} (x-x_0)^2 + b_0 (x-x_0) + a_0\,, \textmd{for} x < x_0\,,
\end{align}
where $c_{\textmd{left}} \leq 0$, the implied variance slope corresponds to $c_{\textmd{left}}$.
For a right extrapolation defined by
\begin{align}
g(x) = c_{\textmd{right}} (x-x_M)^2 + b_M (x-x_M) + a_M\,, \textmd{for} x > x_M\,,
\end{align}
where $c_{\textmd{right}} \geq 0$ and $b_M = g'(x_M)$, the implied variance slope corresponds to $c_{\textmd{right}}$.

\begin{figure}[H]
	\centering{
		\subfigure[Case II of \ref{tbl:jackel} with $\epsilon = 10^{-4}$]{
			\includegraphics[width=.48\textwidth]{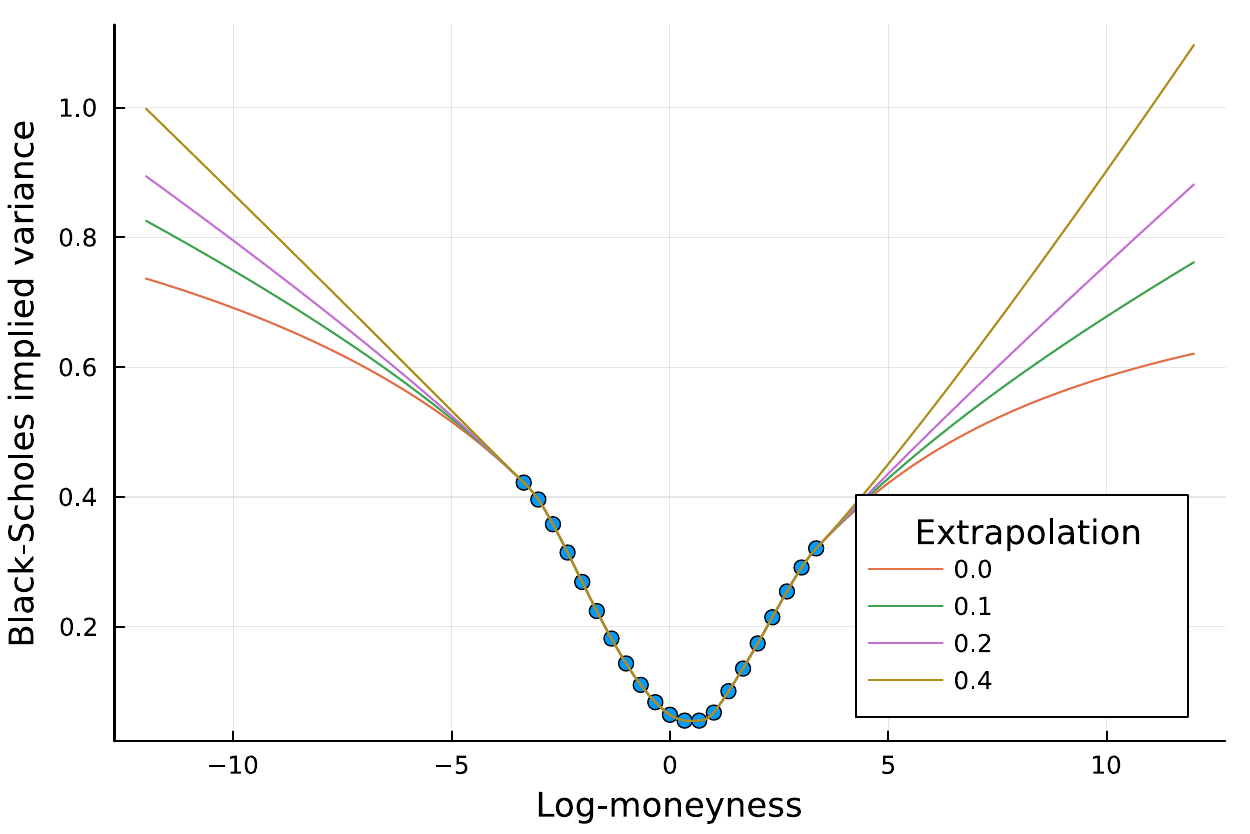}}
		\subfigure[TSLA with $\epsilon=10^{-2}$]{
			\includegraphics[width=.48\textwidth]{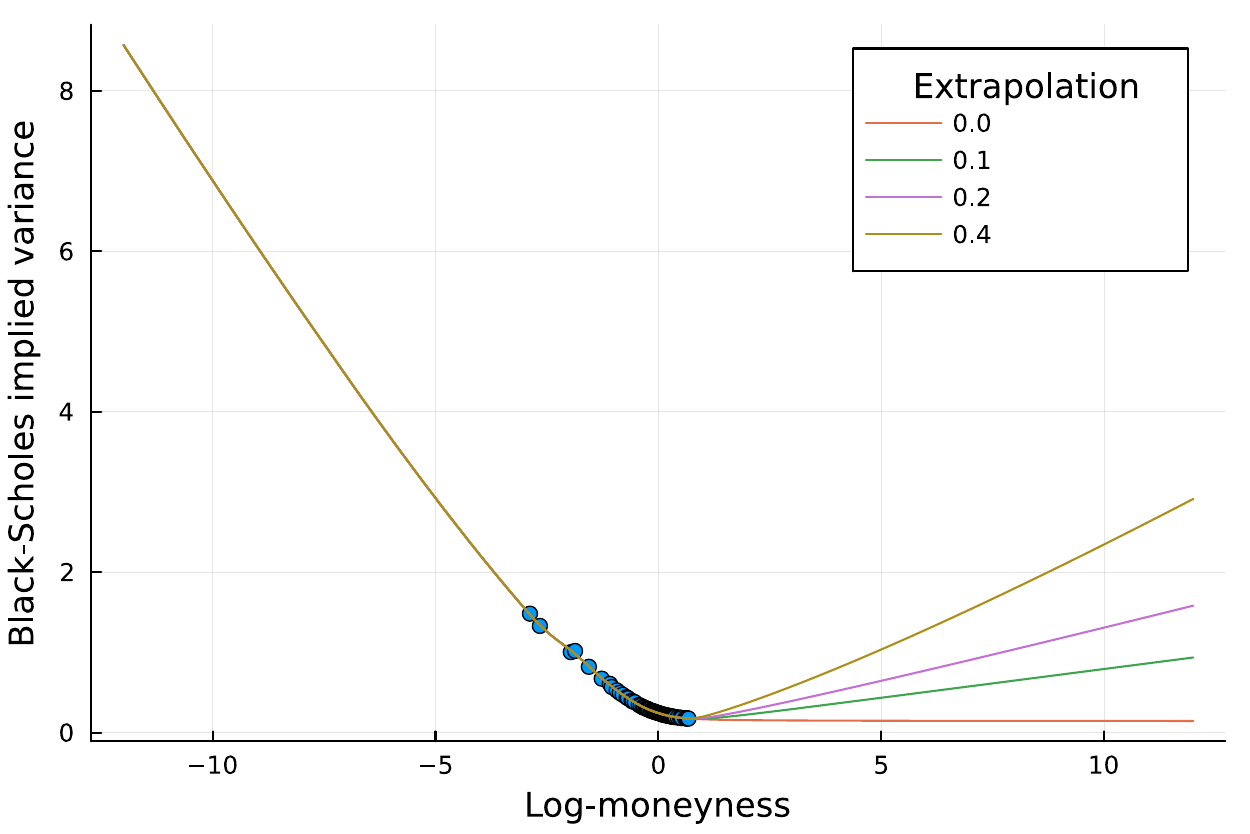} }
	}
	\caption{Implied variance of the exponential B-spline collocation calibrated with different extrapolation curvatures.\label{fig:ive_expbspline_tsla_eps01}}
\end{figure}

During the calibration to vanilla options, we fix the left and right extrapolation curvatures $c_{\textmd{left}}$ and $c_{\textmd{right}}$. On the market data of Table \ref{tbl:jackel} (Case II), the impact of the change of curvature becomes noticeable when the log-moneyness is above 5.0. This is because the maturity is long (5 years). The impact is more visible in the right wing when calibrating to TSLA options of maturity 18 months (Figure \ref{fig:ive_expbspline_tsla_eps01}).

Why doesn't it play a role for the left wing as well? This is because $x_0$ corresponds to a strike below 0.30, that is a log-moneyness below -7.3, which is extremely low and the extrapolation curvature does not impact the calibration at all. Note that if we plot the left wing starting at log-moneyness -12.0, we still would not see a difference: the left asymptotic behavior is theoretical and does not really materialize\footnote{unless we go to silly strikes. A strike of 0.30 is already extremely small for an 18 months option on an asset with forward price 356.73.} on this example. 
Even with the Schumaker spline calibration, where  $x_0$ corresponds then to a strike above 30, above the lowest market strike of 20, we don't see a large difference in the left wing: another cause is the large value of $|b_0|$ on this data.

For the right wing, $x_M$ corresponds to a strike of between 835 and 890 (a log-moneyness of around 0.90), which is much closer to the highest market strike of 700. 

\section{Conclusion}
Optimizing abscissae (resp. B-spline knots) is not necessarily more appropriate than ordinates (resp. B-spline coefficients) as the ordinates are not truly fixed in the former approach due to the enforcement of the first moment condition.

Optimizing abscissae also restricts the technique to use specific monotonic spline interpolations, making higher orders of smoothness more challenging to attain.

In general, for a positive asset, it may be more appropriate to use regular B-spline collocation, instead of the exponential version, and employ an exponential linear or quadratic extrapolation as in \citep{lefloc2021positive} to guarantee positivity. The use of a regular B-spline instead of the exponential of a B-spline leads to smoother implied probability density, and allows to easily increase the order of smoothness by increasing the B-spline degree, while keeping the same analytical formula for pricing European options.

Finally, instead of using a penalty it may be simpler and more efficient to use less knots in the spline representation when the goal is to fit a dense set of market option prices.

Those conclusions should be taken with some care as they are based on a very limited (but relevant) set of example market data.

\funding{This research received no external funding.}
\conflictsofinterest{The authors declare no conflict of interest.}

\externalbibliography{yes}
\bibliography{lefloch_collocation}

\begin{thebibliography}{-------}
\providecommand{\natexlab}[1]{#1}

\bibitem[{Le Floc’h} and Oosterlee(2019)]{lefloch2019model}
{Le Floc’h}, F.; Oosterlee, C.W.
\newblock Model-Free Stochastic Collocation for an Arbitrage-Free Implied Volatility, Part II.
\newblock {\em Risks} {\bf 2019}, {\em 7},~30.

\bibitem[Wolberg and Alfy(2002)]{wolberg2002energy}
Wolberg, G.; Alfy, I.
\newblock An energy-minimization framework for monotonic cubic spline interpolation.
\newblock {\em Journal of Computational and Applied Mathematics} {\bf 2002}, {\em 143},~145--188.

\bibitem[Schumaker(1983)]{schumaker1983shape}
Schumaker, L.I.
\newblock On shape preserving quadratic spline interpolation.
\newblock {\em SIAM Journal on Numerical Analysis} {\bf 1983}, {\em 20},~854--864.

\bibitem[Roos(2025)]{roos2025simple}
Roos, T.
\newblock Simple, Flexible, Analytic, Arbitrage Free Volatility Interpolation.
\newblock {\em Available at SSRN 5215592} {\bf 2025}.

\bibitem[Lam(1990)]{lam1990monotone}
Lam, M.H.
\newblock Monotone and convex quadratic spline interpolation.
\newblock {\em Virginia Journal of Science} {\bf 1990}, {\em 41}.

\bibitem[Grzelak and Oosterlee(2017)]{grzelak2017arbitrage}
Grzelak, L.A.; Oosterlee, C.W.
\newblock From arbitrage to arbitrage-free implied volatilities.
\newblock {\em Journal of Computational Finance} {\bf 2017}, {\em 20},~31--49.

\bibitem[J{\"a}ckel(2014)]{jackel2014clamping}
J{\"a}ckel, P.
\newblock Clamping Down on Arbitrage.
\newblock {\em Wilmott} {\bf 2014}, {\em 2014},~54--69.

\bibitem[{LeFloc'h} and Oosterlee(2021)]{lefloc2021positive}
{LeFloc'h}, F.; Oosterlee, C.W.
\newblock Positive Stochastic Collocation for the Collocated Local Volatility Model.
\newblock {\em arXiv preprint arXiv:2109.02405} {\bf 2021}.

\end{thebibliography}
\appendixtitles{no}
\appendix
\section{Market data}\label{sec:market_data_jackel}

\begin{table}[H]
	\centering{
		\caption{Black-Scholes implied volatilities against moneyness $\frac{x}{X(0)}$ for an option of maturity $T=5.0722$, examples 1 and 2 of \citet{jackel2014clamping}.\label{tbl:jackel}}
		\begin{small}
			\begin{tabular}{rrr}\toprule
				\multicolumn{1}{c}{Moneyness} & \multicolumn{1}{c}{Volatility (Case I)} &  \multicolumn{1}{c}{Volatility (Case II)} \\ \midrule
				0.035123777453185 & 0.642412798191439 & 0.649712512502887 \\ 
				0.049095433048156 & 0.621682849924325 & 0.629372247414191\\ 
				0.068624781300891 & 0.590577891369241 & 0.598339248024188 \\ 
				0.095922580089594 & 0.553137221952525 & 0.560748840467284 \\ 
				0.134078990076508 & 0.511398042127817 & 0.518685454812697 \\ 
				0.18741338653678 & 0.466699250819768 & 0.473512707134552 \\ 
				0.261963320525776 & 0.420225808661573 & 0.426434688827871 \\ 
				0.366167980681693 & 0.373296313420122 & 0.378806875802102 \\ 
				0.511823524787378 & 0.327557513727855& 0.332366264644264 \\ 
				0.715418426368358 & 0.285106482185545 & 0.289407658380454\\ 
				1 & 0.249328882881654 & 0.253751752243855\\ 
				1.39778339939642 & 0.228967051575314 & 0.235378088110653 \\ 
				1.95379843162821 & 0.220857187809035 & 0.235343538571543 \\ 
				2.73098701349666 & 0.218762825294675 & 0.260395028879884 \\ 
				3.81732831143284 & 0.218742183617652 & 0.31735041252779 \\ 
				5.33579814376678 & 0.218432406892364 & 0.368205175099723 \\ 
				7.45829006788743 & 0.217198426268117 & 0.417582432865276 \\ 
				10.4250740447762 & 0.21573928902421 & 0.46323707706565 \\ 
				14.5719954372667 & 0.214619929462215 & 0.504386489988866 \\ 
				20.3684933182917 & 0.2141074555437 & 0.539752566560924 \\ 
				28.4707418310251 & 0.21457985392644  & 0.566370957381163\\ \bottomrule
			\end{tabular}
			\label{tbl:jackel_clamping_1}
	\end{small}}
\end{table}

\begin{table}[!ht]
	\centering{
		\caption{Implied volatilities against strikes $K$ for TSLA options expiring on January 17, 2020, as of June 15, 2018. This corresponds to a maturity $T= 1.59178$ and the forward is $f= 356.73$.}
		\begin{tabular}{
S[table-format=5.0]
S[table-auto-round,table-format=6.5]
S[table-format=5.0]
S[table-auto-round,table-format=6.5]}
		\toprule
		\multicolumn{1}{r}{Strike} & \multicolumn{1}{r}{Implied volatility} & \multicolumn{1}{r}{Strike} & \multicolumn{1}{r}{Implied volatility} \\ \cmidrule(lr){1-2} \cmidrule(lr){3-4}
		20 & 1.21744983334323 & 330 & 0.50711984442627 \\ 
		25 & 1.15297355418723 & 335 & 0.504289607300568 \\ 
		50 & 1.00135129931668 & 340 & 0.501395969703038 \\ 
		55 & 1.00870138714102 & 350 & 0.496189725922157 \\ 
		75 & 0.905591957613594 & 360 & 0.491447823711383 \\ 
		100 & 0.819649926900943 & 370 & 0.485710524333137 \\ 
		120 & 0.779704840770866 & 380 & 0.482098230257581 \\ 
		125 & 0.753927847741657 & 390 & 0.477655148504366 \\ 
		140 & 0.725534998629369 & 400 & 0.4682253137831 \\ 
		150 & 0.703696294602874 & 410 & 0.469126243065069 \\ 
		160 & 0.687090759720296 & 420 & 0.465204974999456 \\ 
		175 & 0.663148945950044 & 430 & 0.462103669314557 \\ 
		180 & 0.654280983914334 & 440 & 0.45969798571593 \\ 
		195 & 0.631004843189498 & 450 & 0.456135600518296 \\ 
		200 & 0.623197951319144 & 460 & 0.454181891398352 \\ 
		210 & 0.615452601430001 & 470 & 0.45154516512584 \\ 
		230 & 0.58662148341447 & 480 & 0.445418855804426 \\ 
		240 & 0.578375148373119 & 490 & 0.445283390706062 \\ 
		250 & 0.562503659012476 & 500 & 0.443037556906725 \\ 
		255 & 0.562553917615043 & 510 & 0.439392127793856 \\ 
		260 & 0.557233168461812 & 520 & 0.441317531074983 \\ 
		270 & 0.548521241773961 & 550 & 0.433632202339099 \\ 
		275 & 0.545613165725652 & 580 & 0.429705382102393 \\ 
		280 & 0.540060895711996 & 590 & 0.428435742375435 \\ 
		285 & 0.538477679227124 & 600 & 0.424107747661981 \\ 
		290 & 0.53252981128395 & 650 & 0.422267272903106 \\ 
		300 & 0.522241064755214 & 670 & 0.420343689285221 \\ 
		310 & 0.5202396738775 & 680 & 0.419341951870164 \\ 
		315 & 0.516841425453668 & 690 & 0.419347323460756 \\ 
		320 & 0.512740549020954 & 700 & 0.417589294204178 \\ 
		325 & 0.510044008755892 & \multicolumn{1}{l}{} & \multicolumn{1}{l}{} \\ \bottomrule
	\end{tabular}
	\label{tbl:tsla_jun1518}}
\end{table}

\end{document}